\documentclass[12pt,preprint]{aastex62}
\usepackage{amsmath}
\usepackage{graphicx}
\usepackage{multirow}
\usepackage{color}
\usepackage{natbib}
\usepackage{morefloats}

\begin{document}

\title{A Northern Ecliptic Survey for Solar System Science}

\author[0000-0003-4365-1455]{Megan E. Schwamb}
\correspondingauthor{Megan E. Schwamb}
\email{mschwamb.astro@gmail.com}
\affiliation{Gemini Observatory, Northern Operations Center, Hilo, HI USA}

\author[0000-0001-8736-236X]{Kathryn Volk}
\affiliation{Lunar and Planetary Laboratory, The University of Arizona, Tucson, USA}

\author[0000-0001-7737-6784]{Hsing~Wen~(Edward)~Lin}
\affiliation{Department of Physics, University of Michigan, Ann Arbor, MI, USA}

\author[0000-0002-6702-7676]{Michael S.~P.~Kelley}
\affiliation{University of Maryland at College Park, College Park, MD, USA}

\author[0000-0003-3257-4490]{Michele T. Bannister}
\affiliation{Astrophysics Research Centre, Queen's University Belfast, Belfast, United Kingdom}

\author[0000-0001-7225-9271]{Henry H. Hsieh} 
\affiliation{Planetary Science Institute, Tucson, AZ, USA}
 
\author[0000-0001-5916-0031]{R. Lynne Jones}
\affiliation{University of Washington, Seattle, WA, USA}

\author[0000-0002-8132-778X]{Michael Mommert}
\affiliation{Lowell Observatory, Flagstaff, AZ, USA}
 
 \author[0000-0001-9328-2905]{Colin Snodgrass} 
 \affiliation{University of Edinburgh, Edinburgh, UK} 
 
\author[0000-0003-1080-9770]{Darin Ragozzine}
\affiliation{Brigham Young University, Provo, UT, USA}

\author[0000-0003-3240-6497]{Steven R. Chesley}
\affiliation{Jet Propulsion Laboratory, CA, USA}

\author[0000-0003-3145-8682]{Scott S. Sheppard}
\affiliation{Department of Terrestrial Magnetism (DTM), Carnegie Institution for Science, Washington, DC, UA}

\author[0000-0003-1996-9252]{Mario Juri\'{c}}
\affiliation{University of Washington, Seattle, WA, USA}

\author[0000-0003-0854-745X]{Marc W. Buie}
\affiliation{Southwest Research Institute, Boulder, CO, USA}

\date{November 30, 2018}

\begin{abstract}
        Making an inventory of the Solar System is one of the four fundamental science requirements for the Large Synoptic Survey Telescope (LSST). 
        The current baseline footprint for LSST's main Wide-Fast-Deep (WFD) Survey observes the sky below 0$^\circ$ declination, which includes only half of the ecliptic plane. 
        Critically, key Solar System populations are asymmetrically distributed on the sky: they will be entirely missed, or only partially mapped, if only the WFD occurs. 
        We propose a Northern Ecliptic Spur (NES) mini survey, observing the northern sky up to +10$^\circ$ ecliptic latitude, to maximize Solar System science with LSST.  
        The mini survey comprises a total area of $\sim$5800 deg$^2$/604 fields, with 255 observations/field over the decade, split between \emph{g,r,} and \emph{z} bands. 
        Our proposed survey will 1) obtain a census of main-belt comets; 2) probe Neptune's past migration history, by exploring the resonant structure of the Kuiper belt and the Neptune Trojan population; 3) explore the origin of Inner Oort cloud objects and place significant constraints  on the existence of a hypothesized planet beyond Neptune; and 4) enable precise predictions of KBO stellar occultations. 
        These high-ranked science goals of the Solar System Science Collaboration are only achievable with this proposed northern survey. 
        
\end{abstract}

\section{White Paper Information}

\begin{enumerate} 
\item {\bf Science Category:} Taking an Inventory of the Solar System
\item {\bf Survey Type Category:} mini survey
\item {\bf Observing Strategy Category:}  An integrated program with science that hinges on the combination of pointing and detailed observing strategy 
\end{enumerate}

\clearpage
\section{Scientific Motivation}\label{sec:motivation}

\begin{figure}[h!]
\begin{center}
	\includegraphics[width=0.89\columnwidth]{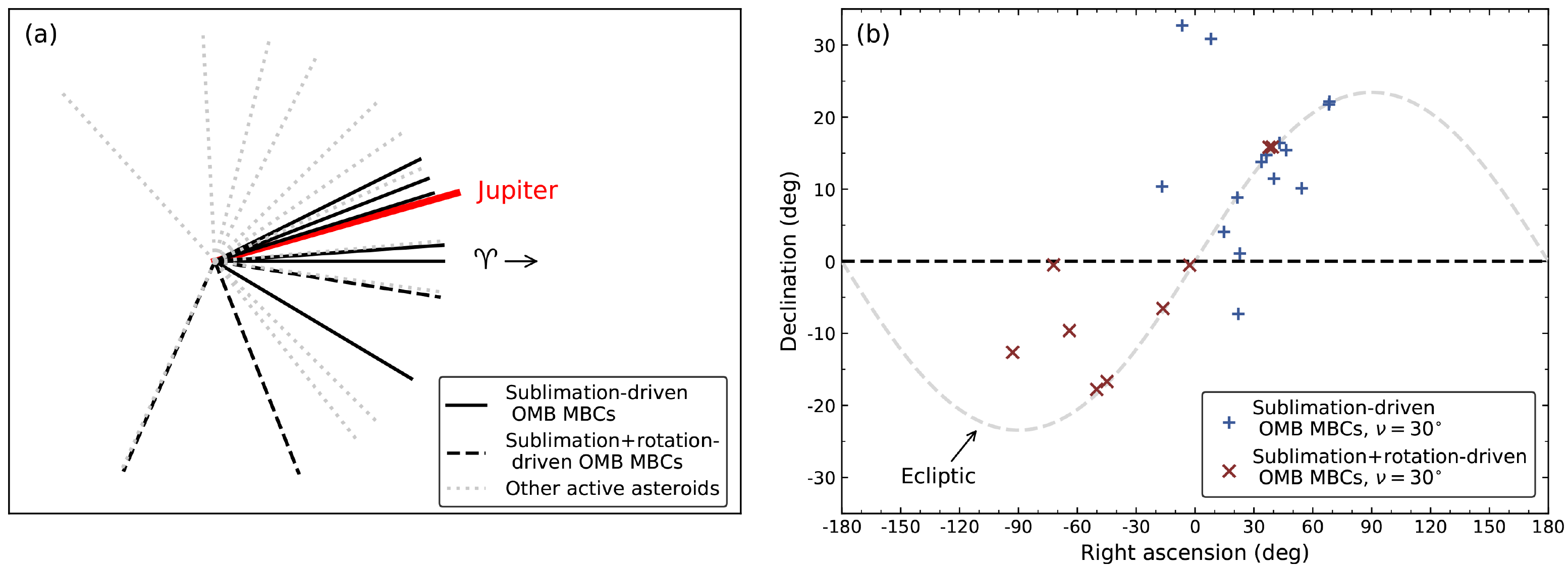}
	\caption{\label{fig:mbcs}\textbf{(a)} Directions of the longitudes of perihelion of outer-main-belt (OMB) MBCs whose activity is attributed to sublimation or a combination of sublimation and rotational destabilization. Adapted from \citet{2018AJ....155..142K}. \textbf{(b)} Sky positions of sublimation-driven OMB MBCs, and sublimation and rotation-driven OMB MBCs when at $\nu=30^{\circ}$ (i.e., when peak activity is expected) over the course of the LSST survey.}
   \end{center}
\end{figure}

\begin{figure}[h!]
\begin{center}
	\includegraphics[width=0.87\columnwidth]{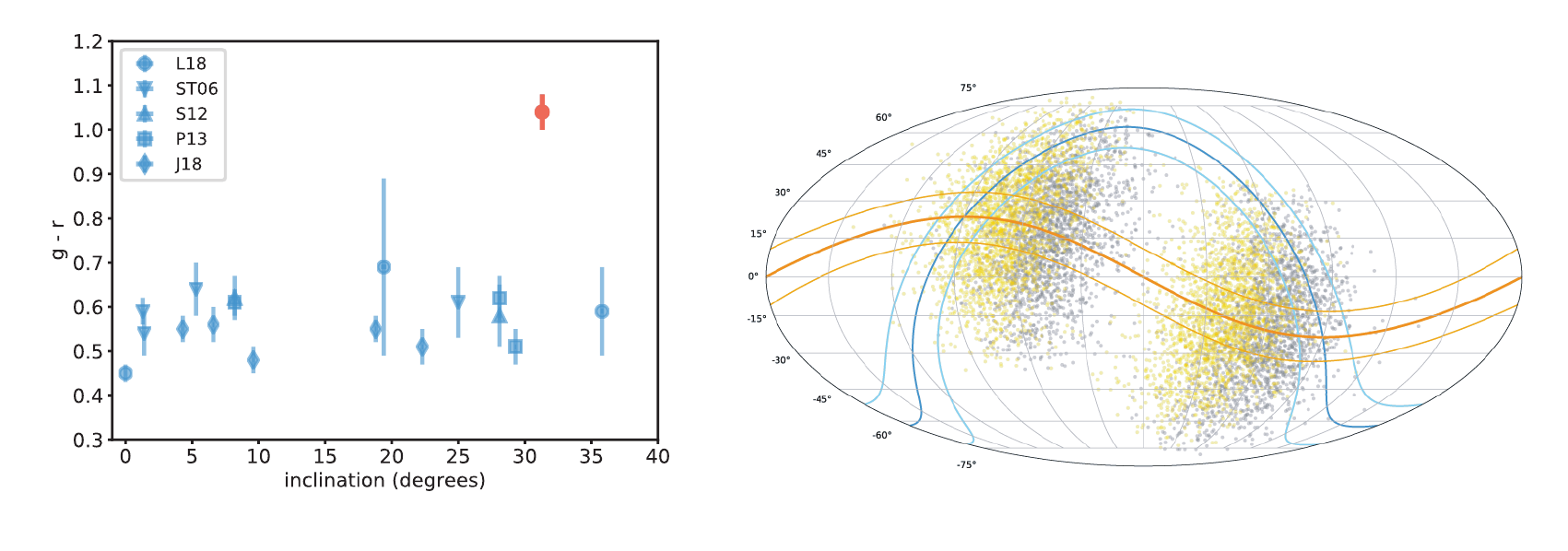}
	\caption{   \label{fig:NT}\textbf{Left:} The color-inclination relation of Neptune Trojans. The only known extra-red Neptune Trojans has second highest inclination of $\sim 31^{\circ}$ (red circle). \textbf{Right:} The on-sky positions of Neptune Trojans in 2022 (grey) and 2032 (yellow).}
	   \end{center}
\end{figure}

\begin{figure}[h!]
\begin{center}
	\includegraphics[width=0.44\columnwidth]{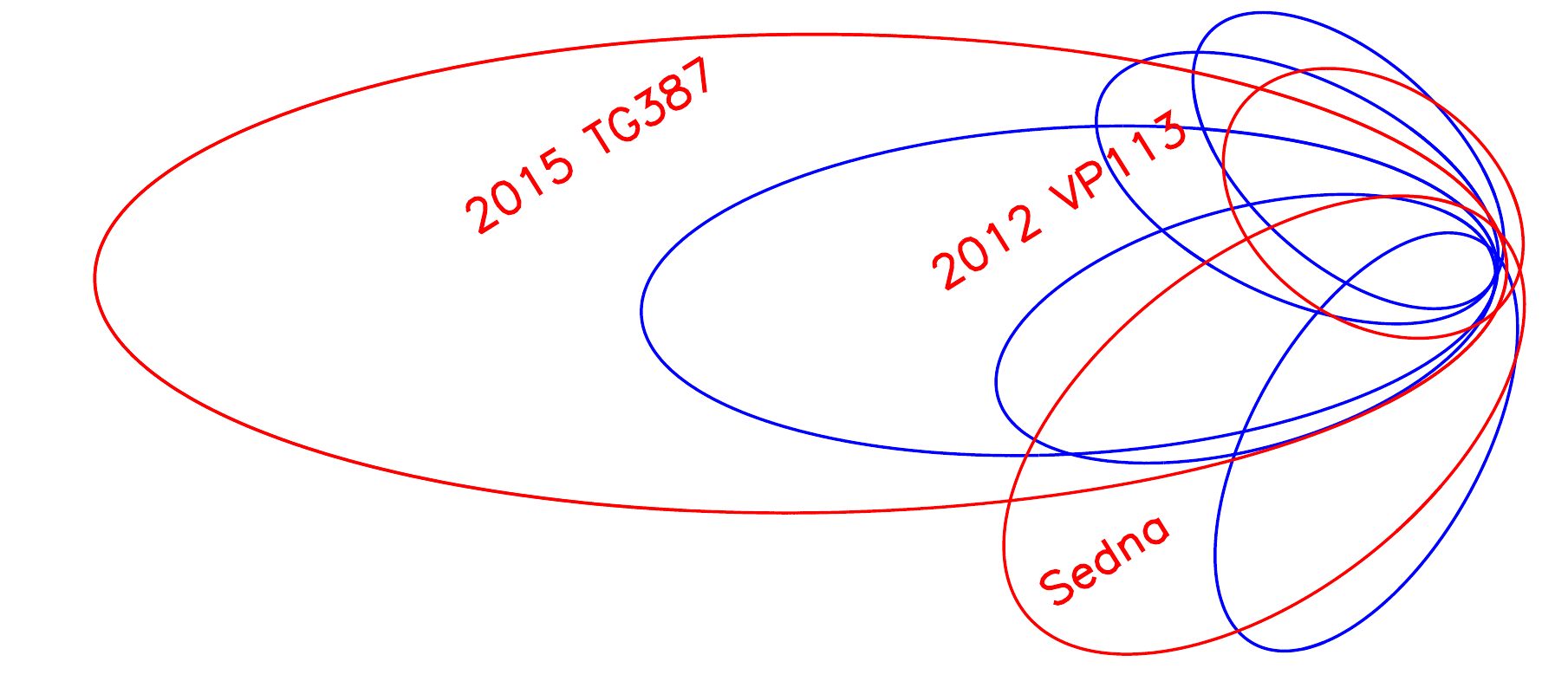}
	\caption{\label{fig:IOCs}  Alignment of known inner Oort cloud orbits, with perihelia beyond 45 au and semi-major axes $250 < a < 2000$~au. Orbits in red have perihelia greater than 60 AU. All these extreme objects come to perihelion within a few hours of 4 hours RA.
	}

   \end{center}
\end{figure}

A foundational goal of the Large Synoptic Survey Telescope (LSST) is to map the Solar System \citep{2008arXiv0805.2366I,2009arXiv0912.0201L}. 
Multiple major small body populations (described below) are key windows into understanding our Solar System's formation and evolution, but are asymmetrically distributed on the sky. 
They will be partially mapped or completely missed without coverage of the Northern ecliptic, which is absent from the Wide-Fast-Deep (WFD) footprint. 
Other yet-unseen asymmetric distributions are likely to exist, and will only be found by surveying the entire ecliptic. To achieve the goals in the LSST Solar System Science Collaboration's roadmap \citep{2018arXiv180201783S}, we propose a Northern Ecliptic Spur (NES) mini survey covering up to $+10^\circ$ ecliptic latitude.
\\
\\
\textbf{Main-Belt Comets:}  Main-belt comets (MBCs) occupy dynamically asteroidal orbits between Mars and Jupiter, yet exhibit comet-like activity near perihelion due to sublimation of volatile ices \citep{2006Sci...312..561H}.  Fewer than a dozen are currently known, where they are considered valuable probes of volatile distribution in the Solar System's primordial disk. They comprise a subset of the active asteroids, which are dynamically asteroidal objects that exhibit activity due to sublimation, rotational destabilization, impacts, and other effects \citep[][]{2015aste.book..221J}. \cite{2018AJ....155..142K} find that almost all of the known MBCs reach perihelion (and thus become active) in the same direction as Jupiter's perihelion, clustering in our proposed NES survey region (Figure \ref{fig:mbcs}). A NES mini survey is needed to (a) determine whether this alignment of MBC perihelia is maintained as more MBCs are discovered, and if this apparent alignment is verified to be real, (b) discover smaller MBCs, which will only be bright enough to detect when near perihelion and active, and (c) monitor known main-belt asteroids for activity at times when they are most likely to become active.
\\
\\
\textbf{The Kuiper Belt's Structure and Neptune's Migration:} The detailed structure of the Kuiper belt, the swarm of planetesimals orbiting beyond Neptune, provides important constraints on early Solar System dynamical history. 
The populations now in mean-motion resonance  constrain Neptune's orbit during its outward migration.
The number of Kuiper belt objects (KBOs) in each resonance constrain high-eccentricity phases and/or semi-major axis jumps during Neptune's migration \citep[e.g.,][]{1995AJ....110..420M,Levison:2008,Nesvorny:2016}.  
Detailed distribution inside within resonances is also valuable. For example, the ratio of KBOs in the leading and trailing libration islands in Neptune's 2:1 resonance, can act as a speedometer for Neptune's migration \citep{Murray-Clay:2005}. KBOs are exceptionally distant, have a steep size distribution, and are thus faint: their discoveries are strongly biased toward detection at perihelion. 
Detectable resonant KBOs are asymmetrically distributed on the sky, as they  come to perihelion at specific geometries relative to Neptune \citep[e.g.,][]{Gladman:2012}. KBOs traverse a minute fraction of their orbit during LSST's 10-year baseline. Surveying the entire ecliptic is critical to observing enough resonant KBOs to make these tests. 
In the absence of the NES, a substantial fraction of key orbital groupings within these important resonances will be completely missed. 
\\
\\
\textbf{L4 Neptune Trojans:} Neptune Trojans co-orbit with Neptune around its L4 and L5 Lagrangian points, emplaced during Neptune's migration. Their orbital/physical property dependencies and L4/L5 population asymmetries are important probes both of Neptune's dynamical history and the Solar System's primordial disk.  \cite{Lin:2018} discovered the first ultra-red  Neptune Trojan, similar to the ultra-red surfaces seen residing within the Kuiper belt (see Figure \ref{fig:NT}). With an inclination $\sim 31^{\circ}$, this discovery may show that ultra-red surfaces only occur at high inclinations, but the origin of this surface type remains unknown. 
\cite{Lin:2016} also find that the larger (H $<$ 8) Neptune Trojans have lower inclinations. Only 19 L4 and 3 L5 Neptune Trojans are known to date; more detections with LSST are needed to confirm these correlations.  Figure \ref{fig:NT} plots the on-sky positions of simulated Neptune Trojans. L5 Neptune Trojans will have good coverage within the WFD survey and the wide inclination distribution of Neptune Trojans make part of the high-inclination L4 Trojans detectable in the WFD. However, the majority of low-inclination L4 Trojans will be missing without the NES. The NES is crucial to test the size-inclination and color-inclination dependencies and  symmetry in properties between the L4 and L5 Neptune Trojans. 
\\
\\
\textbf{Planet 9 and the Origin of the Inner Oort Cloud:} 
Inner Oort Cloud objects (IOCs) are on highly elongated orbits with perihelia beyond 45 au and semi-major axes greater than 250 au and less than 2000 au \citep{2015MNRAS.446.3788B}. IOCs are not significantly influenced by the known inner giant planets or outside forces, but were emplaced by some sort of dynamical interactions, possibly from past stronger outside forces, such as would happen in a dense stellar cluster or from an unseen massive planet \citep{2004ApJ...617..645B}.  As shown in Figure \ref{fig:IOCs}, all of the few known IOCs come to perihelion at similar locations on the sky, which is proposed to be from a distant planet gravitationally shepherding the IOCs onto similar orbits \citep{2014Natur.507..471T,2016AJ....151...22B}.  The IOCs appear to cluster in longitude of perihelia near an RA of $4\pm 3$ hrs, placing most detectable IOCs in the NES fields. 
The NES is thus vital to understand this enigmatic population and determine the true clustering of the IOCs across the sky. If the orbit clustering is real, the NES can expect to find a significant number of IOCs every few hundred square degrees at 24th mag \citep{2016AJ....152..221S}, constraining the orbit of the proposed planet. In addition, \cite{2016ApJ...824L..23B} predict the planet itself would be in the NES fields and detectable by LSST. Alternatively, the current wide range of IOC formation scenarios, e.g. the size of the Sun's birth cluster, will be significantly constrained.
\\
\\
\textbf{KBO Stellar Occultations:} A NES mini survey will recover most of the $\sim$1800 known northern KBOs, providing precise 10-year observation arcs calibrated to Gaia astrometry \citep{2018A&A...616A...2L}. It will be possible to accurately predict stellar occultations of these bodies. Stellar occultations enable accurate measures of sizes, albedos, and binarity, and also probe for rings, atmospheres, and topographic features \citep[e.g.][]{2013ApJ...773...26B,2014Natur.508...72B,2017AJ....154...22D,2017Natur.550..219O}. No planned survey will reach the same  depth as LSST in the NES: the occultation science gains are unique to the NES mini survey.

\section{Technical Description}

\subsection{High-level description}

\begin{figure}[!h]
\begin{center}
	\includegraphics[width=0.7\columnwidth]{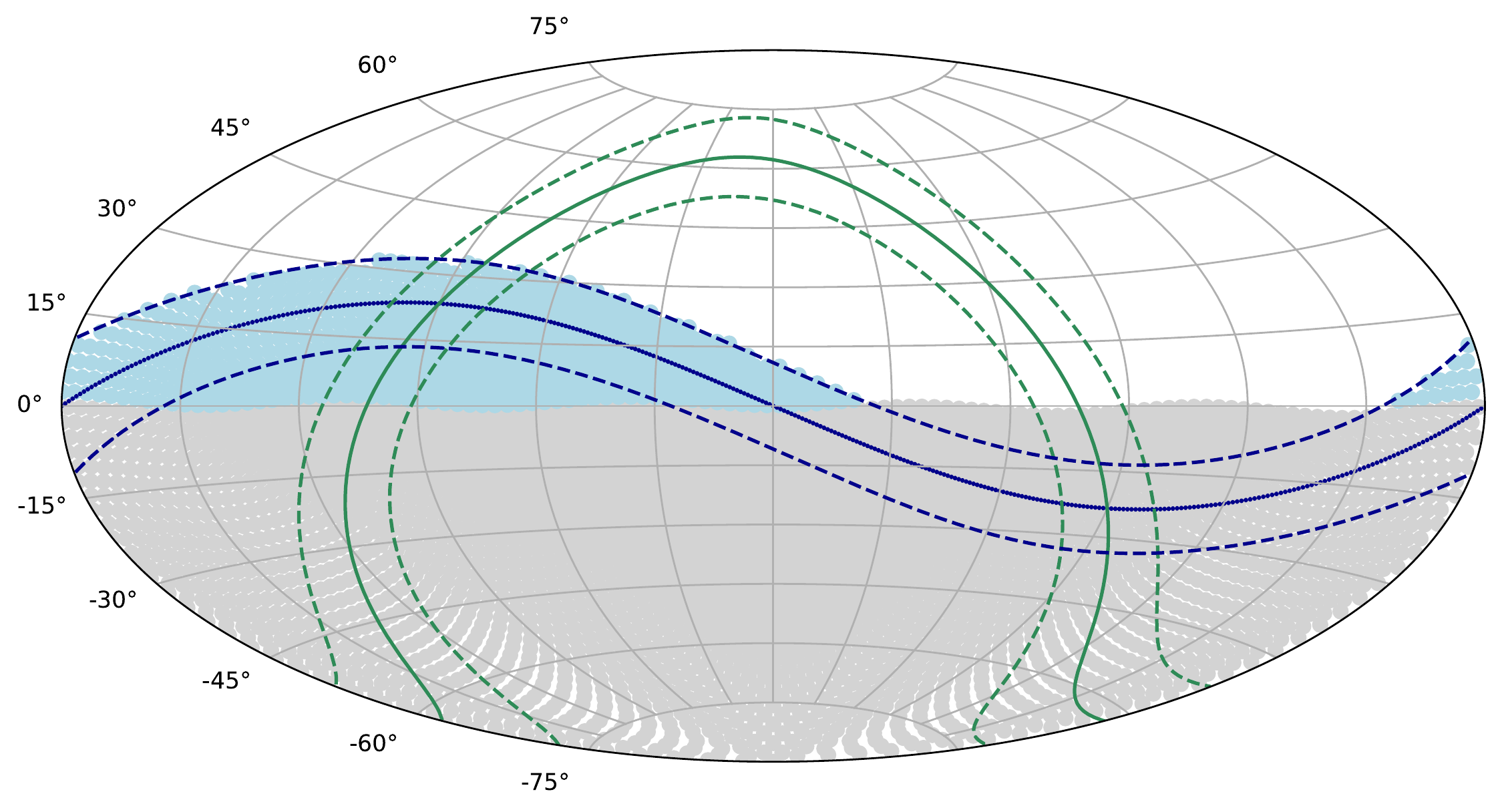}
	\caption{Light blue represents the pointings requested in our proposed Northern Ecliptic Survey. The solid blue line represents the ecliptic. The dashed blue lines represent $\pm$ 10 degrees ecliptic latitude. The solid green line plots the center of the galactic plane. The dashed green lines reflect $\pm$ 10 degrees galactic latitude.}
   \label{fig:footprint}
   \end{center}
\end{figure}

We propose a mini-survey with observations covering the ecliptic plane beyond the region covered within the main Wide-Fast-Deep (WFD) Survey footprint, in order to fully sample small body populations throughout the Solar System. Our proposed NES contains the missing 50\% of the total area of the ecliptic on the sky that is not contained within the WFD survey -- and has about the same fraction of Solar System small bodies at any time.  In the absence of the NES, the substantial fraction of objects near the ecliptic in the Northern hemisphere would be completely missed. We are requesting 255 visits per field over ten years in $grz$, over our mini survey region (the `Northern Ecliptic Spur' region, NES) 
reaching from the northernmost limit of the WFD up to an ecliptic latitude of $+10^{\circ}$ (see Figure~\ref{fig:footprint}).  We assume LSST's discovery and attribution performance will be as described in \cite{2018Icar..303..181J}. The cadence of observations is important, in order to enable linking and tracking; 
we are requesting 6 visits in pairs per night, for each of 5 months in 7 `Discovery' years, 
with 15 visits per year over (split between 5 months) in 3 `Tracking' years;  
this is similar to but dramatically less densely sampled than the WFD baseline strategy. Section \ref{sec:distvisits} describes in detail our preference for how these observations should be divided over the 10-year LSST operational baseline.  The details in this section, including the number of fields and fraction of time required, is based on analysis of a series of simulations created with the LSST Operations Simulator \citep[{\tt OpSim},][]{2014SPIE.9150E..15D} for the call for cadence optimization white papers. 

\clearpage
\subsection{Footprint -- pointings, regions and/or constraints}

In order to complete coverage of the ecliptic plane, we propose a mini survey footprint ranging from the 
northernmost top of the WFD ($\approx \delta = 0^\circ$) up to an ecliptic latitude of $+10^{\circ}$ 
(see Figure~\ref{fig:footprint}). This correlates to 604 distinct LSST fields, using the pointing tessellation
provided in the current simulations. This sky coverage is the best compromise between the northern declination limit of the telescope and the inclination distributions of the Solar System's small body reservoirs. Although orbits of all inclinations cross the ecliptic, those bodies on orbits with higher inclinations spend most of their time away from the ecliptic plane. Additionally some of the KBO resonant populations are perturbed by Neptune such that they come to closest approach off of the ecliptic. By covering fields to 0 degrees declination as the lower extent of the NES, we ensure adequate coverage of all our key Northern Solar System populations.

\subsection{Image quality}

The image quality used in the NES images should be similar to what is used in the WFD. There are no special constraints on image quality beyond what has already been set in the current operation simulations of NES fields and what is needed to achieve our desired individual image depths. In particular, the image quality (seeing) constraints for northern ecliptic observations in the baseline2018a and kraken\_2026 {\tt OpSim} runs are sufficient for our needs. 

\subsection{Individual image depth and/or sky brightness}

The individual image depth is important (and thus its implied constraints on image quality and sky brightness), 
as moving objects must be detectable in individual images. As the goal is to discover small bodies at all 
locations in the ecliptic plane without introducing significant bias, the individual image depths in this mini survey region should be similar to the individual image depths in the WFD footprint. There are no hard cutoffs, but there is  an overall preference for visits to be as deep as possible without sacrificing sky coverage. Past wide-field Solar System surveys have reached a limiting magnitude of $\sim$22nd mag in R \citep[e.g ][]{2010ApJ...720.1691S,2016arXiv161205560C,2018ApJ...855L...6H}. Thus for LSST to make a significant contribution to Solar System Science, the 5-$\sigma$ limiting magnitude per exposure in $r$ and $g$ must be greater than 23rd magnitude with exposure times of 30s or more. We propose the same total exposure time per visit as the WFD (30s per visit), which currently meets our detection goals.  In particular, individual image depths in the northern ecliptic observations in the baseline2018a and kraken\_2026 {\tt OpSim} runs are sufficient for our needs. Specific details on the Solar System Object Differential Completeness Goals that are desired for both the WFD footprint and this proposed NES mini survey are detailed in the Community Observing Strategy Evaluation Paper \citep[COSEP;][]{2017arXiv170804058L}. Additionally, we note that sky brightness can be a particular concern for this region, as the distance to the moon will tend to be small, but is not a constraint in and of itself.

\subsection{Co-added image depth and/or total number of visits}

There are no constraints on the co-added image depth. 
We do have constraints on the total number of visits, due to cadence preferences and requirements for identifying moving objects and characterizing their orbits and physical properties. 
These are discussed below in more detail, but result in a total number of visits on the order
of 255 per field.

\subsection{Number of visits within a night}
At least two visits per night to a field are required to detect and identify moving Solar System objects. At least three separate nights are required to identify and link newly discovered moving objects (see the high-level requirements for LSST project's Moving Object Pipeline System (MOPS) defined 
in LSE-30\footnote{\url{http://ls.st/LSE-30}} and  LDM-156\footnote{\url{http://ls.st/LDM-156}}). 
To guarantee discovery of Solar System bodies at 95$\%$ confidence by MOPS, three tracklets (a pair of images 
in the same night, acquired no more than 90 minutes apart) acquired within 15 days are needed. The most distant objects in the Inner Oort cloud region beyond 200 au, do not move appreciably within a single night, but with two visits per night for each field we can use those the two epochs to weed out the faster/closer moving Solar System bodies in order to optimize and speed up the search algorithms. Thus, we propose at least two visits per night to each field when possible. 

\subsection{Distribution of visits over time}
\label{sec:distvisits}

To balance our key science cases for the NES we spread the `Discovery' and `Tracking' years as described in Table \ref{tab:obs_summary}, below. This cadence allows orbital recovery for the different populations and maximizes temporal coverage for main-belt comet and asteroid collision discovery. An additional requirement during the `Discovery' years is an observation schedule that supports MOPS discovery of main-belt objects.  MOPS requires a pair of visits per night, in at least three separate nights, within 15 days in order to identify and link minor planets (resulting in a 95$\%$ confidence of discovery; LSE-30; LDM-156).  Each pair consists of two visits within the same night, no more than 90 minutes apart.  Once the discovery criteria are satisfied for a given field in the `Discovery' year, the remaining observations can be scheduled more flexibly. During `Tracking' years, the measured orbits of previous detected asteroids found in `Discovery' years will be used to predict the locations of these bodies in the tracking observations to check on cometary activity.

\begin{table}[h]
    \centering
    \begin{tabular}{|c|l|l|}
    \hline
            Year & Observation Type & Summary \\ \hline
             1 & Discovery & 6 observations per field per month for 5 months  \\  \hline
             2  & Discovery  & 6 observations per field per month for 5 months  \\  \hline
            3 &  Discovery & 6 observations per field per month for 5 months  \\  \hline
             4 & Tracking  & 15 observations per field divided over 2-3 months \\  \hline
            5 & Discovery  & 6 observations per field per month for 5 months  \\  \hline
             6 & Discovery  & 6 observations per field per month for 5 months  \\  \hline
            7 & Tracking & 15 observations per field divided over 2-3 months \\  \hline
             8 & Tracking  & 15 observations per field divided over 2-3 months \\  \hline
              9 & Discovery  & 6 observations per field per month for 5 months  \\  \hline
            10 & Discovery  & 6 observations per field per month for 5 months  \\  \hline
    \end{tabular}
     \caption{{\bf Proposed NES Year-by-Year Summary} }
        \label{tab:obs_summary}
    \end{table}

The population of MBCs we have discovered in the past decade are likely the brightest and youngest members.  With LSST's superior sensitivity, we will search for activity around all asteroids, with additional scrutiny for those in the outer main-belt.  For known asteroids, we only require a few observations of MBC-like asteroids near perihelion, rather than the more strict circumstances needed for discovery.  This is readily satisfied in both our `Discovery' and  `Tracking' years as described above. The `Discovery' years are also crucial for the MBCs.  Cometary activity enhances the brightness, and therefore discoverability of MBCs.  However, detectable activity for the currently known MBCs has been limited to brief periods near perihelion, typically only a few months long but sometimes even shorter \citep{2015Icar..248..289H}.  Being outer main belt objects, MBC have orbital periods of $\sim$6 years.  Thus, we request 7 `Discovery' years in the NES to increase the discoverability of MBCs during their brief active periods.  We simulated LSST's discoverability of asteroids with MBC-like orbits based on \cite{2018AJ....155..142K}, and found the rate of discovery during periods of activity is up to 35\% without the NES, and up to 55\% with the NES, corresponding to an enhancement of $\sim$1.6. The distribution of observations during `Tracking' years is flexible, but we nominally request a cadence to maximize temporal coverage to search for MBC activity over the 5 months an NES field is typically observable. We request that `Tracking' year observations consist of ideally nightly pairs of visits made on seven different nights spread out over 5 months centered around opposition for each field, with no specific spacing requirements other than the first and last pairs of visits being spaced by at least two months.  This approach would provide somewhat denser and more uniform time sampling of known objects to search for cometary and other activity. 
 
 The proposed multi-year observations of our proposed NES footprint are also designed to characterize the orbits for Outer Solar System bodies both in the Kuiper belt and Inner Oort cloud. At least two years (two-oppositions) of observations of the NES are required to test the Planet 9 hypothesis and probe the origin of objects like Sedna in the Inner Oort Cloud region.  With a single opposition of observations, we will be unable to distinguish the distant Sedna-like Inner Oort cloud objects from our detections residing within the Kuiper belt. A high-perihelion Inner Oort cloud object and a typical scattered disk KBO near aphelion both provide reasonable Keplerian fits to the observations. Follow-up recovery observations one month or a few months later are crucial in reducing the one-year positional uncertainties, but the two families of orbits only diverge sufficiently after a year. The key to understanding the formation of the Sedna-like Inner Oort cloud objects is through accurate orbits. With firmly characterized orbits that identify whether the distant objects beyond 50 au are at their closest point in their orbit, and therefore detached from Neptune, we can begin to explore the dynamics and structure of the Sedna region. Although two oppositions can separate regular scattered disk KBOs from Sedna-like detached Inner Oort cloud orbits, the semi-major axis uncertainty is sufficiently large that further oppositions as requested in our NES survey are needed to full characterize these distant orbits.   

Observations of the NES at later years are also critical to reducing orbit fit uncertainties enough to allow secure identification of objects in mean motion resonances with Neptune. The distribution of Neptune's resonant population provides a critical test of early Solar System dynamical histories, and securely identifying resonant orbits typically requires 3-$\sigma$ semi-major axis uncertainties below $\sim$1\%. This precision would likely be most easily achieved with moderately dense sampling of the NES over two oppositions (multiple nights for each field at opposition and $\sim2$ months before and after opposition) early in the survey and sparsely sampled follow-up at later oppositions in the survey. Shifting the scheduling of the higher-density sampling years and low-density sampling years during the survey should not significantly affect orbit quality because the most important factor is the overall baseline of observations. 

Our Neptune Trojan science case also benefits from the additional `Discovery' years.  As shown in Figure \ref{fig:NT}, over the 10-year baseline, the core of the L4 Neptune Trojan cloud moves significantly on the sky. In Years 1-3, the bulk of L4 Neptune Trojans will be further than 10 degrees from the galactic plane, making them easier to detect. In later years, they will move within 10 degrees of the galactic plane.  The LSST image subtraction pipeline will have to deal with the significant blending from the crowded fields; likely detection efficiency of MOPS will be lower in these regions of the sky. The outer edges of the cloud that were in the galactic plane in 2022 will move out by 2032, and may have a better chance of discovery in less crowded stellar fields. To also aid in the discoverability of L4 Neptune Trojans, we designed  our proposed NES survey to cluster  `Discovery' years in the first few years and then later in the last 2 years of LSST baseline operations.

 While the exact ordering of our `Discovery' and `Tracking' years for the NES within LSST is flexible for achieving our science goals, we note that having three `Discovery' years at the beginning of LSST will help maximize the use of ground-based facilities in the Northern hemisphere to enhance LSST Solar System science. Having early `Discovery' years will ensure that there are Northern Solar System targets for follow-up observations early in the 10 year LSST baseline. For example, newly discovered MBCs could be targeted for spectroscopic follow-up. Expected IOCs could also be targeted for additional astrometric observations to aid in orbit determination. Preliminary photometric measurements for detected KBOs could also be used to select scientifically compelling targets for  follow-up studies to obtain higher precision lightcurve or color measurements.

\subsection{Filter choice}
\label{sec:filter_selection}

Solar System minor planets, visible from reflected sunlight, are brightest in the mid-optical wavelengths. Inner Solar System objects will move sufficiently over LSST's 10-year operational baseline to be imaged at some point in all 5 filters in the WFD footprint. Over a decade Outer Solar System bodies will not move much in their orbits, such that nearly all of the KBOs imaged in year 1 in the NES survey will remain within the NES footprint in year 10.  Thus, our filter choice maximizes the return on Outer Solar System science. Since the NES fields would receive fewer visits than WFD fields, we prioritize the bulk of observations in $r$, with some $g$ and $z$-band imagery for surface color/composition studies. 

As LSST does not have a wide-band $gri$-type filter as used, e.g., by Pan-STARRS \citep{2016arXiv161205560C}, optimum detection efficiency will be in the $r$ band; we thus prioritize observations in this filter, requesting 60$\%$ of the total observation in each field be taken in $r$. Spectral slope from solar-neutral to solar-reddened can be minimally parameterized with the use of $g$.
However, red surfaces will be very faint in $g$, thus a comparatively substantial 25\% of time must be allocated to obtain sufficient SNR to ensure moderate-quality and (potentially) non-simultaneous $g-r$ colours on LSST's detections.
\citet{Pike:2017} showed that the cold classical population of TNOs display a distinct colour in the colour-space of the filters $g,r,z$.
We thus add $z$ as our third filter for the remaining 15\% of time, in preference to $i$, where no such population distinctiveness is seen.
Most minor planets are very faint reflectors in $u$, so we (reluctantly) omit it from our mini survey. 

Observations within a single night do not necessarily need to be in the same filter, however we will be constrained in detection efficiency by the shallower limiting magnitude of the pair.  Outer Solar System objects have much redder surfaces compared to the inner Solar System bodies. $(g-r)$ colors range to 1 for the very red surfaces in the dynamically excited and cold classical Kuiper belt. To maximize the detection of the reddest KBOs and IOCs, we ideally request that when possible the nightly pairs be taken in $r$. When this is not possible, we request the second nightly visit should be in $g$  since Solar System objects will be faintest in z-band. We request that the various $g,r,z$ nightly  filter pair combinations in a NES mini survey observations with a cadence as proposed here be simulated to better quantify the impact on discovery metrics with an improved KBO SED (spectral energy distribution) in the {\tt OpSim} moving object package.

\subsection{Exposure constraints}

The goal of the NES is to detect sufficient numbers of Northern Solar System objects to characterize the asymmetric distributions of MBCS, resonant KBOs, Neptune Trojans, and IOCs. Ideally, our requirement would be that the proposed NES mini survey is to have the same or better sensitivity to Solar System objects as in the Southern Ecliptic covered in the WFD footprint. This suggests that we aim for a similar NES detection threshold as in the Southern Ecliptic that will be covered by the WFD footprint. There are trade-offs between exposure time and coverage (discussed further below), but our nominal plan uses the same 30s exposure time in the NES as in the WFD survey. Though longer exposures go deeper, the loss in coverage is probably detrimental to our coverage goals which are crucial since these populations are relatively sparse on the sky. Future work could employ metric-driven optimization to investigate these details. 

\subsection{Other constraints}

None noted.               

\subsection{Estimated time requirement}

\begin{itemize}
\item In total 604 fields that have field centers greater than or equal to declination of 0 up to an ecliptic latitude of +10 degrees.
\item 7 `Discovery' year with 30 visits (6 observations per field per month for 5 months)
\item 3 `Tracking' year with 15 observations per field spread out over 2/3 months
\item In total 255 visits for each fields in ten years.
\item Time required per visit is (30 second exposure time + 3 seconds slew/settle + 2 seconds shutter open/close) = 35 second for one snap (see \ref{sec:snaps}). It will be (2 $\times$ 15 second exposure time + 3 seconds slew/settle + 2 $\times$ 2 seconds shutter open/close) = 37 second for two snaps.

\end{itemize}

The total time request will be (604 fields $\times$ 255 visits $\times 35$ second) = 5,390,700 second = 1,497.4 hours for one snap or (604 fields $\times$ 255 visits $\times 37$ second) = 5,698,740 second = 1,583.0 hours for two snaps. This is approximately equivalent to 187 nights total, over the lifetime of LSST, or about 5-6$\%$ of the total available time. Noted that the NES will request 176.2/186.2 hours (one/two snaps) for a `Discovery' observed year and 88.1/93.1 for a `Tracking' year.


\begin{table}[ht]
    \centering
    \begin{tabular}{l|l|l|l}
        \toprule
        Properties & Importance \hspace{.3in} \\
        \hline 
        Image quality &  1   \\
        Sky brightness & 3 \\
        Individual image depth & 1  \\
        Co-added image depth &  3 \\
        Number of exposures in a visit   & 3  \\
        Number of visits (in a night)  &  1 \\ 
        Total number of visits &  2 \\
        Time between visits (in a night) & 1 \\
        Time between visits (between nights)  & 1  \\
        Long-term gaps between visits & 2\\
        Separation between First and Final observation & 2 \\
        Filter Selection & 1 \\
        Number of Snaps in a Visit &  3 \\
       \hline
    \end{tabular}
    \caption{{\bf Constraint Rankings:} Summary of the relative importance of various survey strategy constraints, ranked from 1=very important, 2=somewhat important, 3=not important.}
        \label{tab:obs_constraints}
\end{table}

\subsection{Technical trades}

 \subsubsection{What is the effect of a trade-off between your requested survey footprint (area) and requested co-added depth or number of visits}
 
Trading survey area for co-added depth/number of visits will lead to an increasingly biased sample of Solar System object discoveries or a decreased number of discoveries.  Decreasing the NES survey area will likely decrease our longitudinal or inclination coverage, adversely affecting the observed distributions of objects in the Kuiper belt resonances, Neptune Trojan clouds, MBC reservoirs, and the Inner Oort Cloud.  Increased co-added depth or number of visits per field does not make up for the missing orbital phase space.  In addition, discovery of moving targets requires multiple observations of the same field.  Reducing the number of visits per field in order to increase areal coverage will affect the moving object pipeline's ability to discover unknown moving objects.

 \subsubsection{If not requesting a specific timing of visits, what is the effect of a trade-off between the uniformity of observations and the frequency of observations in time? e.g. a `rolling cadence' increases the frequency of visits during a short time period at the cost of fewer visits the rest of the time, making the overall sampling less uniform.}
 
Our science goals can be carried out with the NES being observed with a higher cadence of observations in 7 years for discovery and orbit characterization with 3 years of sparser monitoring observations in between these discovery/orbit characterization years. There is some flexibility in when the `Tracking' lower number of observation years are scheduled as described in the Sections above. We have some flexibility between the frequency of observations and uniformity of observations, as long as the cadence of the visits are such that during `Discovery' years MOPS is able to successfully run and detect moving objects. MOPS needs three tracklets (a pair in the same night made no more than 90 minutes apart) over 15 days, to guarantee minor planet discovery with 95$\%$ confidence (see LSE-30 and  LDM-156). Once the discovery criteria are satisfied, additional observations should be scheduled in nightly pairs when possible, but the frequency of the observations from year to year can vary. 

\subsubsection{What is the effect of a trade-off on the exposure time and number of visits (e.g., increasing the individual image depth but decreasing the overall number of visits)?}

Increasing the image depth would increase the 5-$\sigma$ limiting magnitude which would increase the number of objects detected, but we require at least 2 observations per night in 3 pairs for MOPS identification in our `Discovery' years. The additional visits proposed are to characterize and confirm the orbit of the new NES discoveries. At least two years of observations are needed to fully secure the outer Solar System orbits.  Decreasing the number of observations will also lower the opportunities to detect main-belt comets. MBCs are mainly expected to be visible when active (i.e., brighter), so decreasing the number of visits will decrease the chances of finding main-belt comets. 
 
 \subsubsection{What is the effect of a trade-off between uniformity in number of visits and co-added depth? }
 
All of our science goals are constrained not by co-added image depth, but by the 5-sigma detection depth in individual LSST frames. There is no significant gain in trading off the uniformity in number of visits for increasing co-added depth. 

\subsubsection{Is there any benefit to real-time exposure time optimization to obtain nearly constant single-visit limiting depth?}
There would be a small benefit for real-time exposure time optimization to obtain a nearly constant single-visit limiting magnitude, as the 5-sigma limiting depth of exposures drives what Solar System objects will be detectable in the proposed NES observations. Given the high airmass ($>$2) that these observations are normally scheduled, the expected benefit from exposure time optimization will be small (see \url{http://astro-lsst-01.astro.washington.edu:8080/allMetricResults?runId=2})

\subsubsection{Are there any other potential trade-offs to consider when attempting to balance this proposal with others which may have similar but slightly different requests?\label{sec:snaps}}

\noindent
\textbf{Snaps:} \\
  Aside from questions of image quality and cosmic ray rejection, which we do not consider here, the SSSC finds very little benefit in having two 15-sec snaps co-added to form one 30 sec visit. Rather, the gain in survey efficiency from eliminating the time lost for the shutter throw and CCD (Charge-Coupled Device) read-out between snaps would be better used for additional observing time during the survey. Furthermore, combining the CCD read with the slew between visits allows for slower read times and thus reduced read noise.

There are two possible benefits to SS science from separate snaps:
\begin{enumerate}
\item Snaps would allow us to ascertain the direction of motion of trailed Solar System detections, which could potentially ease linking to companion trails in the transient stream. However, if there is a companion, it must be near one of two obvious positions, with a known length and orientation. The companion can be found by searching both directions, leading to only a twofold increase in computational effort (for a relatively small number of objects with significant trailing). 

\item A few small near-Earth asteroids rotate rapidly enough that they have photometric variation on the time scale of 15 s, so rotation information could be extracted from two snaps. However, this is only for a small fraction of small objects, and thus represents a tiny fraction of the small body object catalog. Moreover, it is not clear whether the photometric variation from snaps would be sufficient to constrain the rotation period of so-called super-fast rotators.
\end{enumerate}
 Based on the priorities in the SSSC's Science Roadmap \citep{2018arXiv180201783S}, we consider these benefits as minor in comparison to gaining an addition 1-2$\times$10$^5$ additional survey observations which would increase the number of Outer Solar System detections in key populations.  Moving the Wide-Fast-Deep survey to one 30s snap per observation would add 7 percent of the operations time to on-sky observations. The Solar System metrics described in the COSEP \citep{2017arXiv170804058L} and in this white paper will show no negative impact from moving to one 30s snap. Thus, we advocate for the elimination of snaps in order to accommodate observing the Northern Ecliptic Spur and other proposed  mini surveys and deep drilling fields. \newline
 \\
\textbf{Filter Selection:} \\
As described in Section \ref{sec:filter_selection}, $g$,$r$,and $z$-band observations best suite on science cases, with the majority of the proposed observations taken in $r$-band. If the NES is restricted to single-band observations, our minimum discovery needs require $r$-band. We note that observing the NES in filters without  $r$ or $g$ observations would result in significant losses for Solar System detections based on the discovery metrics in the COSEP \citep{2017arXiv170804058L} and the metrics described in Section \ref{sec:metrics}. \newline
\\
\textbf{Extended WFD Footprint:} \\
We have proposed the minimum number of observations and filters that we believe will achieve our key science goals. Increasing the number of filters and increasing number of visits for all or part of the NES in order to accommodate other science cases such as an extended WFD footprint will not negatively impact our science goals, as long as the majority of the 604 fields in the NES region are surveyed. Additional visits would enable better characterization of rotational variability and provide increased sampling in the search for MBCs. \newline
\\
\textbf{Distribution of Observations Within Tracking Years:} \\
We have some flexibility in scheduling observations within Tracking years if there are strong tensions with other proposed observing needs. Instead of evenly distributing nightly over the months the field is observable, for example, Tracking Years could consist of three pairs of visits to a field using the same cadence used during `Discovery' years in the month when a field is at opposition, and one pair of visits per month in the 2 months before and 2 months after opposition.  This approach would preserve some minimal ability to discover newly active objects (which may have been too faint due to the absence of activity to be detectable in previous years) during `Tracking' years, while also continuing to monitor known objects throughout their available observing windows.  \newline
\\
\textbf{A Big Sky Approach to Cadence Diplomacy:} \\
We note that our requested NES footprint is also part of the extended footprint proposed in the `A Big Sky Approach to Cadence Diplomacy" White Paper (Olsen et al) . Our proposed NES mini survey is compatible with the observing scheme outlined in their proposal. We propose that each NES field receive approximately 255 visits over the baseline survey. This value is very similar to the number of visits propose for the NES and other extended footprint regions proposed by Olsen et al. 

\section{Performance Evaluation}

\label{sec:metrics}
We quantify the impact of including and excluding the NES in LSST survey operations. We simulate LSST observations for representative orbital distributions\footnote{Orbital distributions used in our assessments are available via a public GitHub repo \url{https://github.com/lsst-sssc/SSSC_LSST_Cadence_Optimization_Orbit_Test_Populations}.} based on current observational constraints using existing {\tt OpSim} runs and Metric Analysis Framework \citep[MAF; ][]{2014SPIE.9149E..0BJ} moving object tools. We find for our metrics, astro-lsst-01$\_$2039,which does not include the NES region, underperforms.

\begin{figure}[h!]
\begin{center}
	\includegraphics[width=0.92\columnwidth]{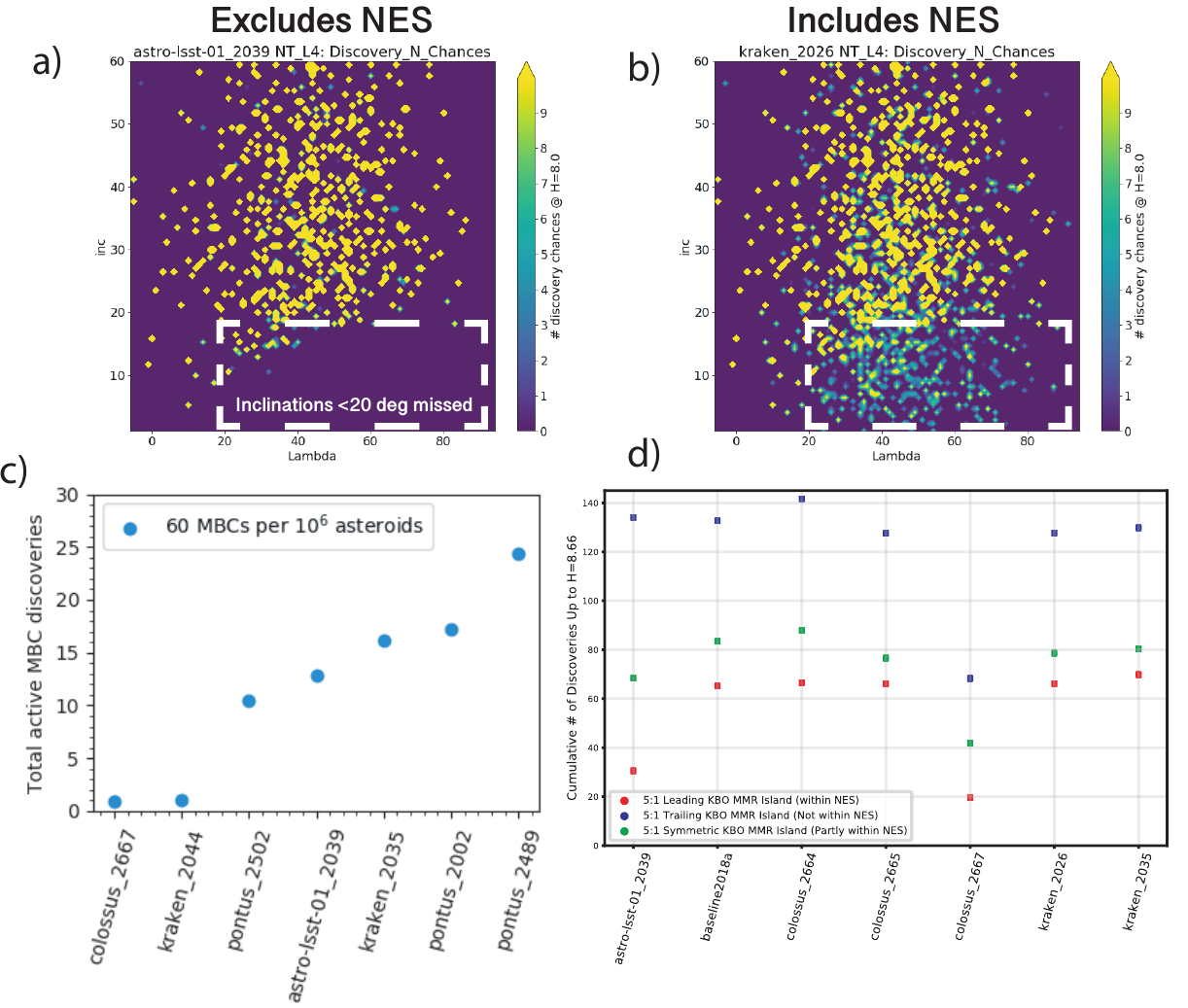}
	\caption{\label{fig:metrics} Metric results for a) Neptune Trojans excluding the NES b)Neptune Trojans including the NES, c) MBCs, d) 5:1 KBO Mean Motion Resonance Libration Islands }
\end{center}
\end{figure}

\textbf{Main-Belt Comets:}  For outer MBCs on eccentric orbits, activity is confined to time periods near perihelion ($r_h\lesssim2.9$~au). %
Our MBC test population has properties consistent with the currently known population 
\citep{2018AJ....155..142K}. 
Using this test population, we propose the following metric to test the discoverability of MBCs:
(1) We take the ad hoc assumption that activity discovery requirements mirror MOPS discovery requirements (6 visits in 15 days; {\tt DiscoveryMetric}); (2) We require discovery circumstances to occur in a 1-month period after perihelion \citep{2015Icar..248..289H} (custom metric).  The $\sim$10 known MBCs are active near perihelion, with a bias to post-perihelion epochs ($-30$ to $+60$ days about perihelion seems typical), which is currently based on a sample of 10 objects.  We base our metric on the period 0 to $+$30 days to allow for some additional diversity in the population. Our metric tested with select available {\tt OpSim} runs is shown in \ref{fig:metrics} (part c)  In summary, excluding the NES reduces the number of MBC discoveries due to the perihelion alignment noted in the Section 2.  \citet{2015Icar..248..289H} estimated an occurrence rate of about 60 MBCs per 10$^6$ outer main belt asteroids.  Taking this rate, the size-frequency distribution of asteroids \citep{2002aste.book...71J}, and the current set of \texttt{opsim} runs, we estimate the number of LSST MBC discoveries to be 10--15 without the NES, and 20--25 with the NES.  The increased number allows us a better estimate of how MBC properties vary,
and assess the lifetime of water ice in $\sim200$~m objects in the outer belt.

\textbf{Resonant KBOs and L4 Neptune Trojans:}

The effect of including the NES on the science return for the outer Solar System can be quantified in terms 1) We need a sufficient number of detections across many different dynamical populations with accurately determined orbits to constrain models of the early Solar System. 2) We need sufficiently accurately determined orbits to classify the detections into these different populations. For each {\tt OpSim} run, the Neptune Trojans and objects in Neptune's 2:1 and 5:1 mean motion resonances are used to determine the expected number of detections for each population. {\tt OpSim} runs with no north ecliptic coverage have extremely few detections for Neptune Trojans in the L4 cloud (none at low inclinations crucial for testing the color-inclination relationship.) and few detections in the leading libration islands of Neptune's N:1 resonances; excluding the north ecliptic cuts the total number of expected discoveries in approximately half (see Fig.~\ref{fig:metrics} a and b). For the case of the 5:1 resonance (Figure 5 d), there would be too few detections to usefully constrain the ratio of leading to trailing populations to test Neptune's migration speed. For distant N:1 resonances, losing half the detections would limit the accuracy of population estimates; we would ideally like the Poisson sample size uncertainties to be less than $\sim15\%$, even for the more distant resonances.

 \textbf{Outer Solar System Orbit Metric:} A metric to measure the orbit fit quality for detected objects will need to be constructed, accounting for the total number of observations assuming an appropriate color distribution. To securely classify objects, we typically need the orbit fit to have a $3-\sigma$ semi-major axis uncertainty $\Delta a/a < 0.01$. This is necessary to separate out resonant and non-resonant objects because Neptune's mean motion resonances are of order $\sim 1$~au wide for typical Kuiper belt eccentricities, though resonances embedded in the classical Kuiper belt can be narrower. 
 
 \textbf{IOCs:} The predicted IOC orbital distribution is dependent on the formation model. Given that the IOCs can only be efficiently discovered near their perihelion and the Planet 9 model predicts perihleion clustering, we suggest a simple metric for success of IOC discovery using fractional coverage of the NES region within 20$^\circ$ of the ecliptic may be most suitable.   

\section{Special Data Processing}

For understanding detection efficiency and characterizing the survey losses, having the same detection algorithm used in the Northern Ecliptic as Southern Ecliptic will be extremely beneficial. In the WFD, MOPS will be the primary moving object search algorithm for Solar System bodies at distances less than approximately 200 au. The proposed observations are designed such that LSST Project-developed MOPS will be able to generate tracklets and link them to identify moving objects in the NES. 

One science case that requires an additional pipeline is the search for very distant small bodies, including additional
Sedna-like objects or Planet 9. MOPS is designed to detect motion between the two visits of the same field within 
a night, separated by $\sim$15-90 minutes. Distant Solar System bodies moving slower than this will be identified as
stationary sources by MOPS. Thus for identifying the most distant Solar System objects beyond a heliocentric 
distance of 200 au and directly detecting Planet 9 (hypothesized to have a semi-major axis of $\sim$700 au), 
a separate detection pipeline will need to be developed by the planetary community. This pipeline task has 
been identified as one of the key tasks in the SSSC Software Roadmap. We note that several members of the 
SSSC have written versions of a slow moving object pipeline \citep[e.g. ][]{2004ApJ...617..645B,2010ApJ...720.1691S,2016AJ....152..221S,2017AJ....153..262B, 2017ApJ...839L..15G,2018ApJ...855L...6H} 
for other outer Solar System surveys and have the expertise to develop such a community pipeline. We also
note that this pipeline could reasonably work on the sources generated from individual images, rather than
requiring the image pixels directly, and can further reject a large majority of the sources in each individual image
immediately as correlated with (long-term) stationary objects; the relevant inputs are relatively small compared to LSST 
data processing.

Main-belt comet science will also require specialized data processing in the form of advanced activity detection 
and characterization software that go beyond the basic activity detection and characterization performed by the 
standard LSST pipelines. These pipelines will build upon the alert stream and LSST produced Solar System data 
products. This specialized software is equally essential for comet science in general for LSST (i.e., including 
observations as part of the main Wide-Fast-Deep survey in the South), and development of this software is already 
a high priority for the SSSC software development and active objects working groups.  No additional special 
data processing requirements beyond what is already planned to be developed to handle comet data from LSST 
in general will be imposed by this mini survey.

\section{Acknowledgements}

 The authors thank the  Large Synoptic Survey Telescope (LSST) Project Science Team and the LSST Corporation for their support of LSST Solar System Science Collaboration's (SSSC) efforts. This work was supported in part by a LSST Corporation Enabling Science grant. The authors also thank the B612 Foundation, AURA, and the Simons Foundation for their support of workshops, hackathons, and sprints that lead to the development of this white paper. Elements of this work were enabled by the Solar System JupyterHub service at the University of Washington's DIRAC Institute (\url{http://dirac.astro.washington.edu}). This white paper has made use of NASA's Astrophysics Data System Bibliographic Services. This version of our NES whitepaper was formatted using the AASTex latex classfile and template package from America Astronomical Society (AAS) Journals \url{http://journals.aas.org/authors/aastex/aasguide.html}.

\bibliographystyle{aasjournal}
\bibliography{references} 
\end{document}